\documentclass[12pt]{article}
\usepackage{graphicx,amssymb,amsmath,empheq,amsthm}
\usepackage{authblk}
\usepackage{comment}
\usepackage{braket}
\usepackage{subfig} 
\usepackage{soul}
\usepackage{hyperref}
\hypersetup{colorlinks = true, linkcolor=black, citecolor=black, urlcolor=blue}
\usepackage{url}

\usepackage[nosort]{cite}

\theoremstyle{plain}

\usepackage{tikz}
\usepackage{tikz-cd}
\usetikzlibrary{arrows}
\usetikzlibrary{intersections}
\usetikzlibrary{shapes.geometric}
\usetikzlibrary{decorations.pathmorphing, patterns,shapes}
\usetikzlibrary{decorations.markings}

\tikzset{
	mid arrow/.style={postaction={decorate,decoration={
				markings,
				mark=at position .575 with {\arrow{stealth}}
	}}},
	near arrow/.style={postaction={decorate,decoration={
				markings,
				mark=at position .275 with {\arrow{stealth}}
	}}},
	far arrow/.style={postaction={decorate,decoration={
				markings,
				mark=at position .800 with {\arrow{stealth}}
	}}},
	snake arrow/.style={fixed point arithmetic, decorate, decoration={snake,amplitude=2pt, segment length=11pt},postaction={decoration={markings,mark=at position 0.625 with {\arrow{stealth}}},decorate}},
}

\tikzset{
  baseline = -0.5ex,
  wavy/.style = {
    thick,
    decorate,
    decoration={snake,amplitude=2pt,segment length=5pt}},
  swavy/.style = {
    thick,
    decorate,
    decoration={snake,amplitude=1.2pt,segment length=3pt}},
  sdot/.style = {
    circle,
    draw=none,
    fill=black,
    minimum size=2.5pt,
    inner sep=0pt},
  bdot/.style = {
    circle,
    draw=none,
    fill=black,
    minimum size=4pt,
    inner sep=0pt},
  svertex/.style = {
    circle,
    draw=black,
    thick,
    fill=lightgray,
    minimum size=8pt,
    inner sep=1pt},
  mmvertex/.style = {
    circle,
    draw=black,
    thick,
    fill=lightgray,
    minimum size=10pt,
    inner sep=1pt},
  mvertex/.style = {
    circle,
    draw=black,
    thick,
    fill=lightgray,
    minimum size=12pt,
    inner sep=1pt},
  bvertex/.style = {
    circle,
    draw=black,
    thick,
    fill=lightgray,
    minimum size=16pt},
  bbvertex/.style = {
    circle,
    draw=black,
    thick,
    fill=lightgray,
    minimum size=20pt}
 }

\renewcommand{\bar}{\overline}
\renewcommand{\tilde}{\widetilde}

\renewcommand{\leq}{\leqslant}
\renewcommand{\geq}{\geqslant}

\newcommand{\Tr}{\operatorname{Tr}}

\newcommand{\R}{{\mathrm{R}}}
\newcommand{\A}{{\mathrm{A}}}

\newcommand{\calP}{\mathcal{P}}

\newcommand{\calS}{\mathcal{S}}

\makeatletter
\def\@fnsymbol#1{\ensuremath{\ifcase#1\or \dagger\or \ddagger\or
   \mathsection\or \mathparagraph\or \|\or **\or \dagger\dagger
   \or \ddagger\ddagger \else\@ctrerr\fi}}
\makeatletter

\numberwithin{equation}{section}

\makeatletter

\newcommand*{\wideboxed}[1]{\setlength{\fboxsep}{1ex}%
  \fbox{\m@th$\displaystyle#1$}}
\makeatother

\makeatletter
\newcommand{\colim@}[2]{%
  \vtop{\m@th\ialign{##\cr
    \hfil$#1\operator@font colim$\hfil\cr
    \noalign{\nointerlineskip\kern1.5\ex@}#2\cr
    \noalign{\nointerlineskip\kern-\ex@}\cr}}%
}
\newcommand{\colim}{%
  \mathop{\mathpalette\colim@{}}\nmlimits@
}
\makeatother

\newcommand{\VF}{\Upsilon}

\renewcommand{\leq}{\leqslant}
\renewcommand{\geq}{\geqslant}

\title{Operator Size Distribution in Large $N$ Quantum Mechanics of Majorana Fermions
}

\author[1,2]{Pengfei Zhang}
\author[3,2]{Yingfei Gu\thanks{guyingfei@gmail.com}}

\affil[1]{\normalsize\it Department of Physics, Fudan University, Shanghai, 200438, P.R.C.}
\affil[2]{\normalsize\it California Institute of Technology, Pasadena, CA 91125, U.S.A.}
\affil[3]{\normalsize\it Institute for Advanced Study, Tsinghua University, Beijing, 100084, P.R.C.}

\date{\today}

\begin{document}
\maketitle

\begin{abstract}
    Under the Heisenberg evolution in chaotic quantum systems, initially simple operators evolve into complicated ones and ultimately cover the whole operator space. We study the growth of the operator ``size'' in this process, which is related to the out-of-time-order correlator (OTOC).   
    We derive the full time evolution of the size distribution in large $N$ quantum mechanics of Majorana fermions. 
    As examples, we apply the formalism to the Brownian SYK model (infinite temperature) and the large $q$ SYK model (finite temperature). 
\end{abstract}

\bigskip

\section{Introduction}

	Quantum scrambling is a process that local information disperses into the entire system under unitary evolution \cite{Hayden:2007cs,Sekino:2008he}. It is vital for the thermalization in isolated quantum systems \cite{PhysRevA.43.2046,PhysRevE.50.888}. 
	Recent developments of information scrambling have connected different branches of physics, including the  quantum theory of gravity \cite{Shenker:2013pqa,Shenker:2014cwa}, non-Fermi liquids in condensed matter physics  \cite{Chowdhury:2021qpy} and many-body quantum chaos \cite{Maldacena:2015waa}. 
	

	In the Heisenberg picture, scrambling manifests itself as the growth of simple operators \cite{Roberts:2014isa}. For a system of $N$ Majorana fermions (with the convention\footnote{The convention here differs from the one commonly used in the SYK related literature by a factor of 2. Here we requires the normalization $\chi_j^2=1$ for the orthonormality of the operator basis $\{ \chi_j,  \chi_{j_1}  \chi_{j_2}, ... ,   \chi_{j_1} \chi_{j_2}... \chi_{j_n}, ... \}$ where the inner product of two operators $X$ and $Y$ is defined as $2^{-N/2} \Tr \small( X^\dagger Y \small) $. The prefactor is the inverse of the  dimension of the Hilbert space.}
	$\{ \chi_j, \chi_k\}=2\delta_{jk}$), 
	a single operator $ \chi_k$ evolves into a superposition of products of Majorana operators under time evolution 
	\begin{equation}\label{eq_expansion}
     \chi_k (t)=e^{i Ht} \chi_k e^{-i Ht} = \sum_{{\rm odd }~n}~ \sum_{j_1< j_2< ...<j_n} c_{j_1 j_2...j_n}(t) ~ \chi_{j_1} \chi_{j_2}... \chi_{j_n},
	\end{equation}
    where the coefficients $c_{j_1j_2...j_n}(t)$ are the (time dependent) amplitudes on the orthonormal operator basis and may be interpreted as the ``operator wave function'' \cite{Roberts:2018mnp}. The length $n$ of the operator string $ \chi_{j_1} \chi_{j_2}... \chi_{j_n}$ is called the {\it ``size''} of this basis element. Accordingly, the probability distribution of the size is given as follows,  
    \begin{equation}\label{distr}
        P(n,t) = \sum_{j_1< j_2< ...<j_n}  \left\vert c_{j_1 j_2...j_n}(t) \right\vert^2.
    \end{equation}
    The conservation of the total probability $\sum_n P(n,t)=2^{-N/2} \Tr\small( \chi_k(t)^\dagger  \chi_k(t)\small)=1$ follows immediately from the unitarity of the time evolution. 
    In this measure, the operator growth is reflected as the shifting of the distribution towards a larger size. 
    
    The notion of operator size is closely related to the out-of-time-order correlator (OTOC)  \cite{larkin1969quasiclassical,kitaev2014talk} in many-body quantum chaos and the momentum of a particle falling towards a black hole horizon in holography\cite{Susskind:2018tei,Brown:2019hmk,Nezami:2021yaq}. 
    However, it is challenging to obtain the analytic expression for operator size and its distribution. Progress has been made in quantum circuits models  \cite{Nahum:2017yvy,Hunter-Jones:2018otn,vonKeyserlingk:2017dyr,Khemani:2017nda,Dias:2021ncd,PhysRevResearch.3.L032057}, numerical studies and experimental proposals for spin models \cite{Roberts:2014isa,Qi:2019rpi,Zhou:2021syv,Omanakuttan:2022ikz}, 
    and bounds on the operator size in various contexts \cite{PhysRevLett.122.216601,Lucas:2020pgj,Chen:2020bmq, Chen:2019klo,Yin:2020oze}. 
    For the large $N$ quantum mechanics like the SYK model \cite{kitaev2015simple,SY}, 
    analytic results were only achieved 
    for the initial growth \cite{Roberts:2018mnp,Qi:2018bje,Lensky:2020ubw}, leaving the saturation in the late time unexplained.  
    In this paper, we  develop a theoretical framework to determine the distribution of size over the full range of time. 
    We apply the framework to two examples: first, the Brownian SYK model where we can compare the analytic result with numerics reported in Ref.~\cite{Sunderhauf:2019djv} and find an agreement to good precision; second, the large-$q$ SYK model at finite temperature.

    \section{Relation to OTOC}

    An operator $O$ in a system of $N$ Majorana fermions ($\chi_j$, $j=1,2...,N$) 
    can be mapped to a state $|O\rangle$ in the doubled system consists of $N$ complex fermions $c_j= (\chi_j + i \psi_j)/{2}$, where $\psi_j$ labels  auxiliary Majorana operators. 
    The mapping can be constructed by acting the operator $O$ on a reference state in the doubled system, e.g.,  $O \mapsto |O\rangle = O |{\rm EPR}\rangle$ with 
    the reference state chosen to be a maximally entangled state between the original system $\chi$ and the auxiliary system $\psi$.

    There are still plenty of freedom in such a maximally entangled state. For the purpose of this paper, we make a particular choice: we require $n_j | {\rm EPR } \rangle=0$ (with $n_j=c^\dagger_j c_j$) for all $j$, namely, $|{\rm EPR}\rangle$ a zero-occupation state for all $N$ flavors.  
    The merit of this choice is that the size of an operator can now be identified with the total fermion occupation number of the corresponding state \cite{Qi:2018bje}. To see this, consider a basis element $O=\chi_{j_1} \chi_{j_2}... \chi_{j_n}$ in the operator space of length $n$, then its corresponding state has the following form 
    \begin{equation}
        |\chi_{j_1} \chi_{j_2}... \chi_{j_n} \rangle = \chi_{j_1} \chi_{j_2}... \chi_{j_n} | {\rm EPR} \rangle =   c^\dagger_{j_1} c^\dagger_{j_2}... c^\dagger_{j_n} | {\rm EPR} \rangle
    \end{equation}
    	as ${\chi}_j={c}_j+{c}^\dagger_j$ and the operator ${c}_j$ annihilates $| {\rm EPR} \rangle$ by construction. For such state, the operator size is equal to the total particle number ${n}:=\sum_j {n}_j$. Consequently, the size distribution \eqref{distr} of the Heisenberg operator $\chi_k(t)$ can be written as the following expectation value 
    	\begin{equation}
    	\label{eq: discrete P}
    	    P(n,t) = \langle {\rm EPR} |  \chi_k^\dagger (t) {\Pi} (n) {\chi}_k (t) | {\rm EPR} \rangle
    	\end{equation}
    	    where ${\Pi}(n)=\delta_{n,\sum_j {n}_j}$ is the projection to the eigenspace with fixed particle number.


In the large $N$ limit, 
instead of the discrete distribution $P(n,t)$ whose argument $n$ is unbounded, 
it is more convenient to use the {\it continuous} distribution $\calP(s,t)$ with the continuous (rescaled) variable $s \in [0,1]$ representing the operator size in unit of $N$:
\begin{equation}\label{calPst}
    	    \calP(s,t) =  \langle {\rm EPR} |  \chi_k (t) {\Pi} (s) {\chi}_k (t) | {\rm EPR} \rangle, 
\end{equation}
where the projection ${\Pi}(s)=\delta(s- \Sigma_j {n}_j/N)$ is a delta function instead of the Kronecker delta. Here, we have dropped the dagger since  $\chi^\dagger_k (t)=\chi_k(t)$ for Majorana operators. 
The discrete and the continuous distribution are related at finite $N$ via equation $\calP(s,t) = N P(sN,t)$, where the prefactor $N$  ensures the normalization $\int_0^1 \mathcal{P}(s,t) ds=1$.

To compute $\calP(s,t)$ in the large $N$ limit, it is useful to consider the following generating function 
       \begin{equation}\label{calS}
       \mathcal{S}(\nu,t) \coloneqq \int_0^1  \mathcal{P}(s,t)e^{-s\nu} ds= \langle \text{EPR}| \chi_k(t) e^{-\frac{\nu}{N} \sum_j  n_j}  \chi_k(t)|\text{EPR}\rangle.
        \end{equation}
The expectation value on the r.h.s. can be mapped back to a Keldysh correlation function in the original system (i.e. only involves $\chi$ fermion) as    
\begin{equation}\label{calSPI}
\begin{tikzpicture}[scale=0.5,baseline={([yshift=-7pt]current bounding box.center)}]
\draw [->,>=stealth] (-50pt,-57pt) -- (180pt,-57pt) node[right]{\scriptsize $t$};
\draw [->,>=stealth] (0pt, -95pt) -- (0pt,-10pt);
\draw[thick,gray] (0pt,-40pt)--(0pt,-50pt);

\draw[thick,gray,far arrow] (0pt,-50pt)--(140pt,-50pt);
\draw[thick,gray,far arrow] (140pt,-54pt)--(0pt,-54pt);
\draw[thick,gray] (0pt,-54pt)--(0pt,-60pt);
\filldraw (140pt,-52pt) circle (2pt);
\node at (150pt,-49pt){\scriptsize$k$};
\draw[thick,gray,far arrow] (0pt,-60pt)--(140pt,-60pt);
\draw[thick,gray,far arrow] (140pt,-64pt)--(0pt,-64pt);
\draw[thick,gray,->,>=stealth] (0pt,-64pt)--(0pt,-80pt);
\filldraw (140pt,-62pt) circle (2pt);
\node at (150pt,-65pt){\scriptsize$k$};
\filldraw[blue] (0pt,-57pt) circle (2pt); \node[blue] at (-12pt,-60pt){\scriptsize $j$};
\filldraw[red] (0pt,-47pt) circle (2pt); \node[red] at (-12pt,-44pt){\scriptsize $j$};
\end{tikzpicture} \qquad
    \mathcal{S}(\nu,t)= \langle {\bf\rm T}_c~ \chi_k(t) e^{-\frac{\nu}{N} \sum_j \frac{1-\chi_j (0) \chi_j(0)}{2} }  \chi_k(t) \rangle,
\end{equation}
where ${\bf\rm T}_c$ denotes the contour ordering (gray line) on the double Keldysh contour and the 
auxiliary fermion fields $\psi_j(0)$ have been mapped to $i\chi_j(0)$ using ${\psi}_j |{\rm EPR}\rangle = i \chi_j |{\rm EPR} \rangle$.

\begin{figure}[t]\centering
\begin{tabular}{c@{\hspace{2.5cm}}c}
\begin{tikzpicture}[scale=1.2]
\node[bvertex] (R) at (-30pt,0pt) {};
\node[bvertex] (A) at (30pt,0pt) {};
\draw[thick] (R) -- ++(135:20pt) node[left]{$1$};
\draw[thick] (R) -- ++(-135:20pt) node[left]{$2$};
\draw[thick] (A) -- ++(45:20pt) node[right]{$3$};
\draw[thick] (A) -- ++(-45:20pt) node[right]{$4$};
\draw[wavy] (A) to[out=140,in=40] (R);
\draw[wavy] (A) to (R);
\draw[wavy] (A) to[out=-140,in=-40] (R);
\end{tikzpicture}
&
\begin{tikzpicture}[scale=1.2]
\node[bbvertex] (R) at (-0pt,0pt) {};
\node[svertex] (A1) at (40pt,24pt) {};
\node[svertex] (A3) at (40pt,-20pt) {};

\draw[thick] (R) -- ++(135:30pt) node[left]{$1$};
\draw[thick] (R) -- ++(-135:30pt) node[left]{$2$};
\draw[thick] (A1) -- ++(70:10pt) node[right]{$3$};
\draw[thick] (A1) -- ++(-20:10pt) node[right]{$4$};
\draw[thick] (A3) -- ++(20:10pt) node[right]{$3$};
\draw[thick] (A3) -- ++(-70:10pt) node[right]{$4$};

\draw[wavy] (R) to[out=60,in=180] (A1);
\draw[wavy] (R) to[out=10,in=230] (A1);
\draw[wavy] (R) to (A3);
\end{tikzpicture}
\vspace{3pt}\\
(a) Diagram for OTOC & (b) Diagram for $\calS(\nu,t)$
\end{tabular}
\caption{(a) and (b) are typical diagrams for OTOC and $\calS(\nu,t)$ respectively.
Wavy lines represent the scrambling modes with propagator $\lambda= C^{-1} e^{\varkappa t}$. At late time $t\sim t_{\rm scrambling}$ when $\lambda \sim 1$, we need to sum over diagrams with multi scrambling modes (wavy lines). 
}
\label{figDiagram}
\end{figure}
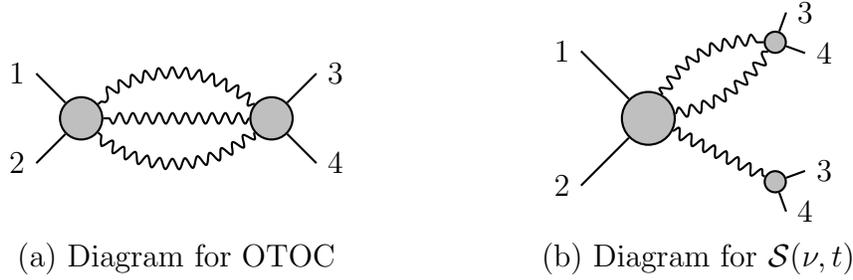

The discussions up to now are generally applicable to quantum systems of Majorana fermions.  Next, we will focus on the interacting systems that exhibit many-body chaos with an exponentially growing OTOC in the early time. This assumption amounts to the following ansatz \cite{Gu:2018jsv, Gu:2021xaj} (this is a non-linear generalization of the ansatz in Ref.~\cite{Kitaev:2017awl})
\begin{equation}
   - \frac{1}{N}\sum_{j} \langle {\chi}_j (t_1) {\chi}_k (t_3) {\chi}_j (t_2) {\chi}_k (t_4) \rangle = \sum_{m=0}^\infty \frac{(-\lambda)^m}{m!} \VF^{\R,m} (t_{12}) \VF^{\A,m} (t_{34})
\end{equation}
where $\lambda = C^{-1} e^{\varkappa \frac{t_1+t_2-t_3-t_4}{2}}$
is regarded as the propagator of the scrambling modes. The prefactor $C$ is of order $N$ and therefore suppresses the higher order terms in OTOC at early time. However, when the center of mass time difference $\frac{t_1+t_2-t_3-t_4}{2}$ is of order the scrambling time $t_{\rm scrambling} \sim \ln N$, 
the exponential growing factor is comparable with $N$. And we need to include the whole series to recover the correct late time behavior, e.g., the saturation of OTOC. Here, the integer ``$m$'' counts the number of scrambling modes (see Fig.~\ref{figDiagram} (a) for an example with $m=3$), and the growing exponent $\varkappa$ in the propagator $\lambda$ is known as the Lyapunov exponent for the many-body quantum system. 
The retarded and advanced vertex functions $(\VF^{\R,m},\VF^{\A,m})$ capture the interactions between the scrambling modes (wavy lines) and the fermions (straight lines, external legs): 
\begin{equation}
\VF^{\R,m}(t_{12})=
    \begin{tikzpicture}
\node[mvertex] (R) at (-20pt,0pt) {};
\node[] (D1) at (0pt,10pt) {};
\node[] (D2) at (0pt,0pt) {};
\node[] (D3) at (0pt,-10pt) {};
\draw[thick] (R) -- ++(135:15pt) node[left]{\scriptsize$1$};
\draw[thick] (R) -- ++(-135:15pt) node[left]{\scriptsize$2$};
\draw[swavy] (D1) to[out=180,in=40] (R);
\draw[swavy] (D2) to (R);
\draw[swavy] (D3) to[out=-180,in=-40] (R);
\end{tikzpicture}
\hspace{20pt}
\VF^{\A,m}(t_{34})=
    \begin{tikzpicture}
\node[mvertex] (R) at (20pt,0pt) {};
\node[] (D1) at (0pt,10pt) {};
\node[] (D2) at (0pt,0pt) {};
\node[] (D3) at (0pt,-10pt) {};
\draw[thick] (R) -- ++(45:15pt) node[right]{\scriptsize$3$};
\draw[thick] (R) -- ++(-45:15pt) node[right]{\scriptsize$4$};
\draw[swavy] (D1) to[out=0,in=140] (R);
\draw[swavy] (D2) to (R);
\draw[swavy] (D3) to[out=0,in=-140] (R);
\end{tikzpicture}.
\end{equation}

Now, for the generating function $\calS(\nu,t)$ of the size distribution expressed as \eqref{calSPI}, its Taylor expansion in parameter $\nu$ consists of OTOCs of multiple operators
\begin{equation}
    \calS(\nu,t) = e^{-\frac{\nu}{2}}  \Bigg\langle {\rm T}_c~ \chi_k(t) \sum_{n=0}^{\infty} \frac{\left( \frac{\nu  }{2N} \sum_j \chi_j(0)\chi_j(0)  \right)^n}{n!}  \chi_k(t) \Bigg\rangle. 
\end{equation}
The prefactor $e^{-\frac{\nu}{2}}$ comes from the constant piece in the exponent in \eqref{calSPI}. 
A typical connected diagram for $\calS(\nu,t)$ is shown in Fig.~\ref{figDiagram} (b) with $n=2$. 
Now, we sum all such diagrams and obtain the following formula in terms of vertex functions and propagators of scrambling modes, 
\begin{equation}
\label{eqnCalS}
    \calS(\nu,t)= e^{-\frac{\nu}{2}}\sum_{n=0}^{\infty} \frac{(\nu/2)^n}{n!} \sum_{m_j\geq 0} \frac{(-\lambda)^{\sum_j m_j}  \VF^{\R,\sum_j m_j} \VF^{\A,m_1}\ldots \VF^{\A,m_n} }{m_1! m_2! \ldots m_n! } ,
\end{equation}
where the time arguments for the vertex functions are zero and have been left out. 

\section{General formula}

Following the manipulations in Ref.~\cite{Gu:2021xaj}, the vertex functions $\VF^{\R/\A,m}$ can be rewritten as the moments of a distribution\footnote{In holography, $h^{\R/\A}$ is related to the distribution of null momentum in single particle states.}
$h^{\R/\A}$, i.e. $\VF^{\R/\A,m}=\int_0^{+\infty}  y^m h^{\R/\A}(y) dy$.
Inserting into \eqref{eqnCalS}, we obtain a simpler expression for the generating function 
\begin{equation}\label{CalSint}
    \calS(\nu,t) =e^{-\frac{\nu}{2}} \int_0^{+\infty} h^\R(y) e^{\frac{\nu}{2} f^\A(\lambda y)} dy,
\end{equation}
whose time dependence is only through the propagator $\lambda = C^{-1} e^{\varkappa t}$. The function $f^{\A}$ in the exponent is the Laplace transform of  $h^{\A}$, namely,  
\begin{equation}
\begin{aligned}
f^\A(\lambda y) =\int_0^{+\infty} e^{-\lambda y z} h^{\A} (z) ~dz=  \sum_{m=0}^{\infty} \frac{(- \lambda y)^m}{m!} \VF^{\A,m} = \begin{tikzpicture}
\draw[thick] (0pt,-10pt) node[right]{\scriptsize$4$}
to[out=135,in=-135] (0pt,10pt) node[right]{\scriptsize$3$};
\end{tikzpicture}\;
+
\begin{tikzpicture}
\node[svertex] (R) at (20pt,0pt) {};
\node[sdot] (D) at (0pt,0pt) {};
\draw[thick] (R) -- ++(45:15pt) node[right]{\scriptsize$3$};
\draw[thick] (R) -- ++(-45:15pt) node[right]{\scriptsize$4$};
\draw[swavy] (D) to (R);
\end{tikzpicture}\;
+
\begin{tikzpicture}
\node[mmvertex] (R) at (20pt,0pt) {};
\node[sdot] (D1) at (0pt,8pt) {};
\node[sdot] (D2) at (0pt,-8pt) {};
\draw[thick] (R) -- ++(45:15pt) node[right]{\scriptsize$3$};
\draw[thick] (R) -- ++(-45:15pt) node[right]{\scriptsize$4$};
\draw[swavy] (D1) to[out=0,in=150] (R);
\draw[swavy] (D2) to[out=0,in=-150] (R);
\end{tikzpicture}\;
+\cdots
\end{aligned}
\end{equation}
It has the meaning of a correlation function on the perturbed background characterized by a mean-field parameter $y$. This function is also relevant in the discussion of quantum teleportation by traversable wormholes \cite{Gao:2019nyj,newpaper}.

The above expression for the generating function 
\eqref{CalSint} 
leads to an integral formula for the size distribution by the inverse Laplace transform
\begin{equation}\label{calPint}
\wideboxed{
    \calP(s,t) = \int_0^{+\infty} h^\R(y) \delta \left( s - \frac{1 - f^{\A}(\lambda y)}{2}  \right)  dy.}
\end{equation}
In other words, the size distribution is connected to the distribution $h^{\R}(y)$ by a change of variable from $y$ to $s=(1-f^\A(\lambda y))/2$. 
The integral with the delta function results in an alternative expression using inverse function
\begin{equation}
    \calP(s,t) = \frac{2h^\R(y)}{ |\partial_y f^\A (\lambda y)|} \qquad \text{with} \quad y= \frac{{f^{\A}}^{-1}(1-2s)}{\lambda}
\end{equation}

The general form of the size distribution function \eqref{calPint}
is our main result, it 
implies the following universal behavior for large $N$ quantum mechanics of Majorana fermions
\begin{enumerate}
    \item In the early time, i.e. when $\lambda = C^{-1} e^{\varkappa t} \ll 1$, we keep to the linear order of $\lambda$ and therefore the size variable $s=\frac{1-f^\A(\lambda y)}{2} \approx \frac{1}{2} \lambda y \VF^{\A,1}$ grows exponentially with time. More precisely, 
the expectation value to its first order approximation 
\begin{equation}
    \overline{s} = \int_0^{+\infty} s~ \calP(s,t)  ds  \approx \frac{1}{2} \int_0^{\infty} \lambda y  \VF^{\A,1} h^\R(y) dy= \frac{1}{2} \lambda \VF^{\A,1} \VF^{\R,1} 
\end{equation} 
grows exponentially as the linearized OTOC, consistent with the result in literature, e.g. Ref.~\cite{Roberts:2018mnp}. 
    \item The novelty of our general result is that it also captures the saturation at late time! For example, at very late time $\lambda \gg 1$, the function $f^\A \ll 1$ and we have 
\begin{equation}
    \calP(s,t) \approx \int_0^{+\infty} h^\R(y) \delta \Big(s-\frac{1}{2}\Big) dy \approx \delta \Big(s-\frac{1}{2} \Big)
\end{equation}
converge to a delta function with $s=1/2$. It means that the initially simple operator is maximally scrambled
after a long enough evolution 
such that a typical component in the linear superposition \eqref{eq_expansion} 
is an operator string 
of length about half of the total flavors, as it is 
generated by random decisions of whether or not to include a particular fermion operator. 
\end{enumerate}

In the following, we will insert known forms of the functions $h^\R$ and $f^\A$ from two SYK-like models
to our general formulas, and show the analytical expressions for the size distribution. 


\section{Brownian SYK and comparison with numerics} 
\begin{figure}[t]
\centering
\begin{tabular}{c@{\hspace{1.5cm}} c}
    \includegraphics[width=0.4\linewidth]{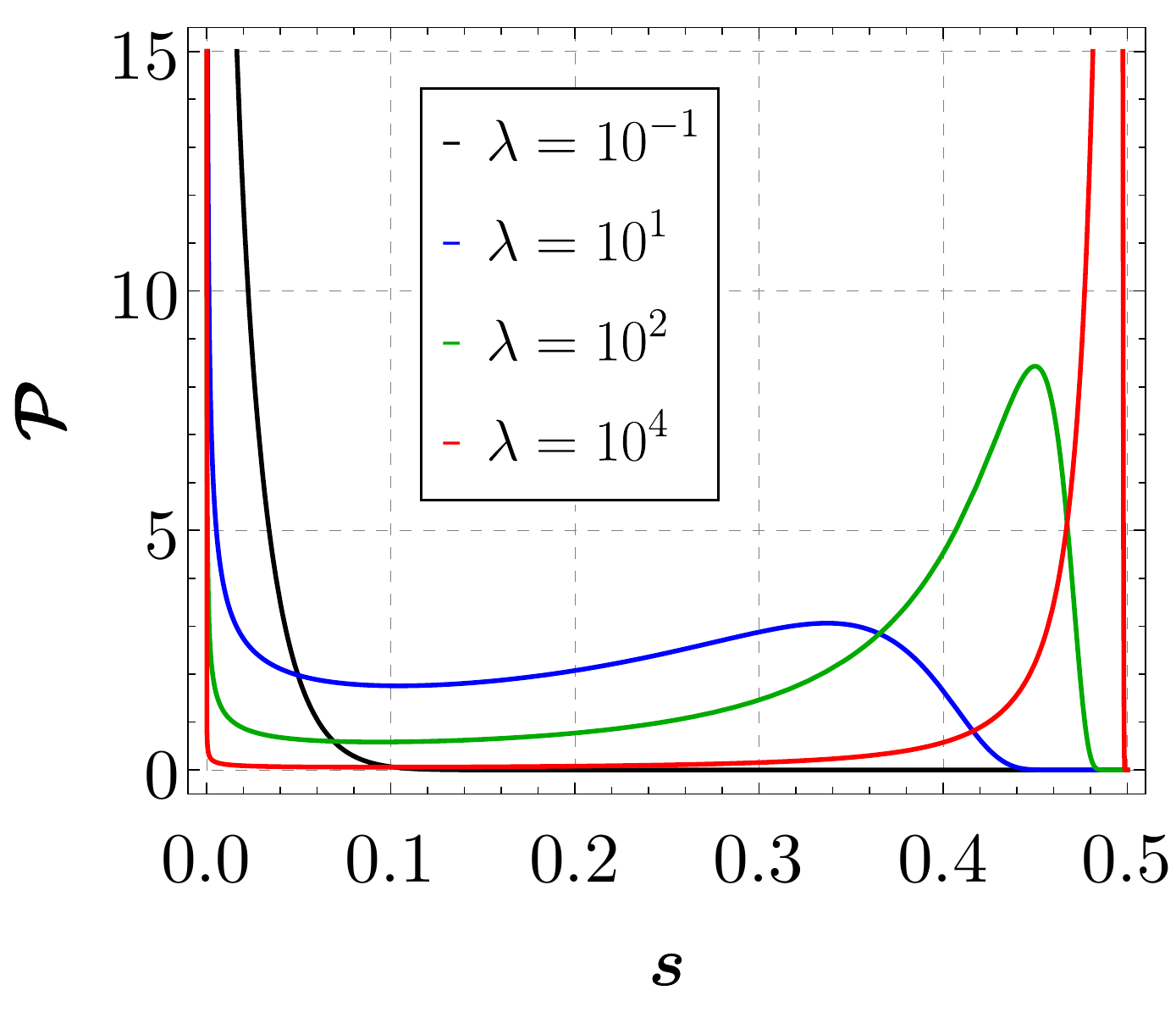} & \includegraphics[width=0.4\linewidth]{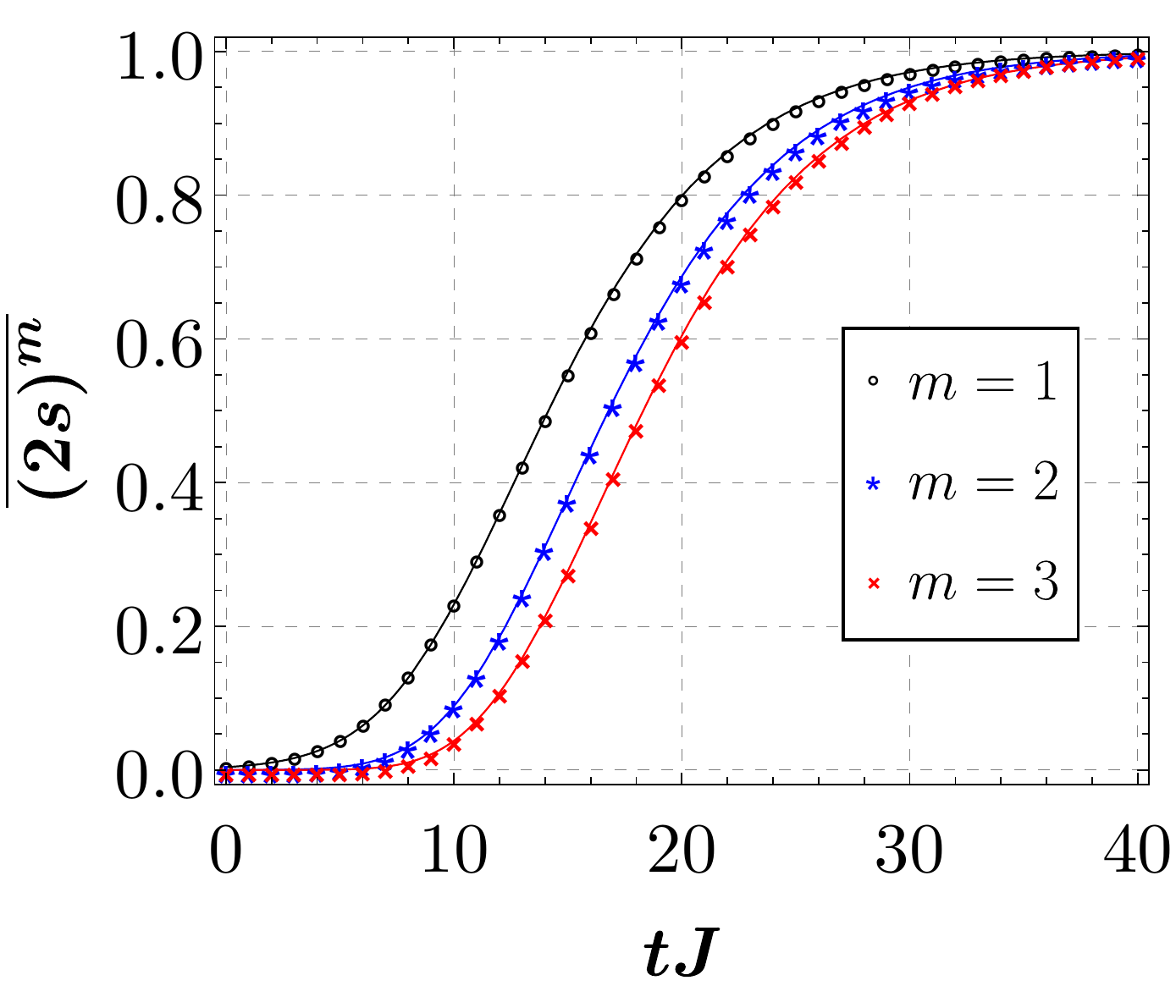} \\
    (a) Size distribution & (b) Prediction vs. Numerics
\end{tabular}
\caption{(a) Size distribution of the Brownian SYK model at different times, parameterized by the values of $\lambda=C^{-1} e^{\varkappa t}$. As time passes (from smaller $\lambda$ to larger ones), the distribution shifts towards larger $s$ and finally accumulates at $s=1/2$;  (b) Comparison between theoretical predictions (solid lines) and numerical results (markers) in the Brownian SYK model with $q=4$ and $N=500$. We plot the moments $\overline{(2s)^m}$ instead of $\overline{s^m}$ so that they saturate to the same value.}
\label{figBrownian}
\end{figure}

The Hamiltonian of the Brownian SYK model \cite{Saad:2018bqo} consists of random all to all time-dependent interactions given as follows
\begin{equation}
     H(t) = i^{\frac{q}{2}} \sum_{1\leq j_1<j_2...<j_q\leq N} J_{j_1j_2...j_q}(t)  \chi_{j_1}  \chi_{j_2} ...  \chi_{j_q}
\end{equation}
where $J_{j_1j_2...j_q}(t)$ are Gaussian variables with zero mean and variance $\overline{J_{j_1j_2...j_q}(t)J_{j_1j_2...j_q}(t')}={(q-1)!J}\delta(t-t')/N^{q-1}$. The functions $h^{\R}$ and $f^\A$ are obtained in \cite{Gu:2021xaj} 
\begin{equation}
    h^\R(y) = \frac{y^{2\Delta-1} e^{-y}}{\Gamma(2\Delta)}, \quad f^\A(x) = \frac{1}{(1+x)^{2\Delta}},~~ \text{with}~\Delta=\frac{1}{2(q-2)}.
\end{equation}
In addition, for the propagator $\lambda = C^{-1} e^{\varkappa t}$, we have the prefactor $C=2N \Delta^2$ and Lyapunov exponent $\varkappa= 4(q-2)J$. 
Putting all these into \eqref{calPint}, we have\footnote{Similar result has been obtained by Shunyu Yao using a different approach \cite{Shunyu}.}
\begin{equation}
\label{eq: brownianP}
    \calP(s,t) = \frac{2 y^{2\Delta-1} e^{-y}}{\lambda \Gamma(2\Delta+1) (1-2s)^{\frac{2\Delta+1}{2\Delta  }}}, \quad y= \frac{(1-2s)^{-\frac{1}{2\Delta}}-1}{\lambda}.
\end{equation}
We plot the size distribution $\calP(s,t)$ in Fig.~\ref{figBrownian} (a), showing a shift of weights towards the larger size and final convergence to a delta function at $s=1/2$.

From \eqref{eq: brownianP} we can also deduce the moments of the size distribution
\begin{equation}
\begin{aligned}
 \overline{s^m} &= \int_0^{+\infty} s^m \calP(s,t) ~ds = \int_0^{+\infty} h^\R(y) \left( \frac{1-f^\A(\lambda y)}{2} \right)^m dy \\
 &= \frac{ 1+\lambda^{-2\Delta}\sum_{k=1}^m(-1)^k {m \choose k} U\left(2\Delta ,1-2(m-1) \Delta ,\lambda^{-1}\right) }{2^m},
\end{aligned}
\end{equation}
where $U(a,b,z)$ is the confluent hypergeometric function. This formula generalizes the results obtained in Ref.~\cite{Stanford:2021bhl}, which are only for the first moment $\overline{s}$. 

Moreover, the average dynamics of the Brownian SYK model can be efficiently simulated using bosonic collective modes, where the relevant Hilbert space dimension grows polynomially with $N$ \cite{Sunderhauf:2019djv} (consequently, the authors were able to study OTOCs upto $N=10^6$). 
We extend the analysis to numerically evaluate the moments $\overline{s^m}$ for $N=500$ and $q=4$, see Appendix~\ref{app: brownian} for details. We compare them with the predictions from our general results in Fig.~\ref{figBrownian}~(b) and find good agreement to high precision.

\section{Finite temperature operator size and the large-$q$ SYK}

So far, our definition and discussions of the operator size have been limited to the infinite temperature system as we were focusing on the Heisenberg evolution of operators rather than the underlying states/ensembles (e.g. the EPR state that was used in \eqref{calPst} is a purification for an infinite temperature density matrix). 
It is also of great interest to generalize the notion of operator size to finite temperatures. However, there is no consensus yet on how to take into account the renormalization effect from the thermal ensemble. Existing proposals include \cite{Qi:2018bje,Lucas:2018wsc}. 
In this paper, we will propose a definition related to the operator ``$Q$'' in Ref.~\cite{Gu:2021xaj} (cf. section 4.3), which measures the ``magnitude of the TFD deformation''. 

Recall that in infinite temperature case, we have mapped an operator $O$ to a state in the doubled system  $| O \rangle = O | {\rm EPR} \rangle$, and related the operator size of $O$ to the fermion number in $|O \rangle$. In this situation, $|{\rm EPR} \rangle$ (i.e. the infinite temperature TFD) is regarded as the vacuum that the fermion operator ${c}_j=({\chi}_j+i{\psi}_j)/2$ annihilates, and the size counts how many excitations an operator creates from the vacuum. 
Now, for finite temperatures, the vacuum in the doubled system is identified as the TFD $|\sqrt{\rho_\beta}\rangle$ which purifies the thermal density matrix. 
We map an operator $O$ to a symmetric\footnote{There are two natural ways to apply an operator to the TFD, which give $|O\sqrt{\rho_\beta}\rangle$ and $|\sqrt{\rho_\beta} O\rangle$. Here we choose a symmetric definition.} ``excited state'' 
$|O\rangle = | \rho_\beta^{1/4} O \rho_\beta^{1/4} \rangle $ and define the operator size as a new type of fermion number in the state  
\begin{equation}
\label{eq: defSize}
    \text{size}[O]_\beta =\frac{ \langle O| n_\beta | O\rangle}{\langle O | O \rangle },   \qquad n_\beta = \sum_j \xi_j^\dagger \xi_j. 
\end{equation}
Here, the fermion operator $\xi_j$ is designed to annihilate the TFD state. $c_\beta = \langle \chi_j (i{\beta}/{4} )  \chi_j (-i{\beta}/{4}  ) \rangle = ( \cos \frac{v\pi}{2} )^{2\Delta}$ is a normalization factor. We use the following definition \cite{Gu:2021xaj}
\begin{equation}
\label{eq: xi}
    \xi_j = \frac{ \chi_j (i\frac{\beta}{4})+ i  \psi_j(-i\frac{\beta}{4}) }{2\sqrt{c_\beta} }, \quad \xi_j^\dagger = \frac{ \chi_j (-i\frac{\beta}{4})- i  \psi_j(+i\frac{\beta}{4}) }{2\sqrt{c_\beta}}
\end{equation}
with the convention 
$ \chi_j (t+i\tau) = e^{-H\tau}  \chi_j(t) e^{H\tau}$ for the imaginary arguments. 

Next, we apply the definition \eqref{eq: defSize} for the finite temperature operator size and calculate the size distribution for a Heisenberg evolved simple fermion operator $\chi_k(t)$ in the large-$q$ SYK model. 
The model was introduced by Maldacena and Stanford in Ref.~\cite{Maldacena:2016hyu} and is analytically tractable at all temperatures. The Hamiltonian is given as follows
\begin{equation}
    H= i^{\frac{q}{2}} \sum_{1\leq j_1<j_2...<j_q\leq N} J_{j_1j_2...j_q}  \chi_{j_1}  \chi_{j_2} ...  \chi_{j_q},
\end{equation}
where $J_{j_1j_2...j_q}$ are random Gaussian variables with zero mean and variance $\overline{J_{j_1j_2...j_q}^2}=\frac{(q-1)!\mathcal{J}^2}{2qN^{q-1}}$. By definition, $N$ is taken to infinity before $q$ goes to infinity. We keep $\mathcal{J}$ fixed in this limit.  A more convenient pair of parameters for this model are $\Delta=1/q$ and $v$, the latter is related to the coupling through equation $v=\frac{\beta \mathcal{J}}{\pi} \cos \frac{\pi v}{2}$. 
$\Delta$ is the fermion scaling dimension, and $v$ determines the time scale in this model, e.g., the Lyapunov exponent $\varkappa=\frac{2\pi v}{\beta}$. The normalization factor $c_\beta =  ( \cos \frac{v\pi}{2} )^{2\Delta}$ for this model. 

\begin{figure}[t]
    \centering
    \includegraphics[width=0.4\linewidth]{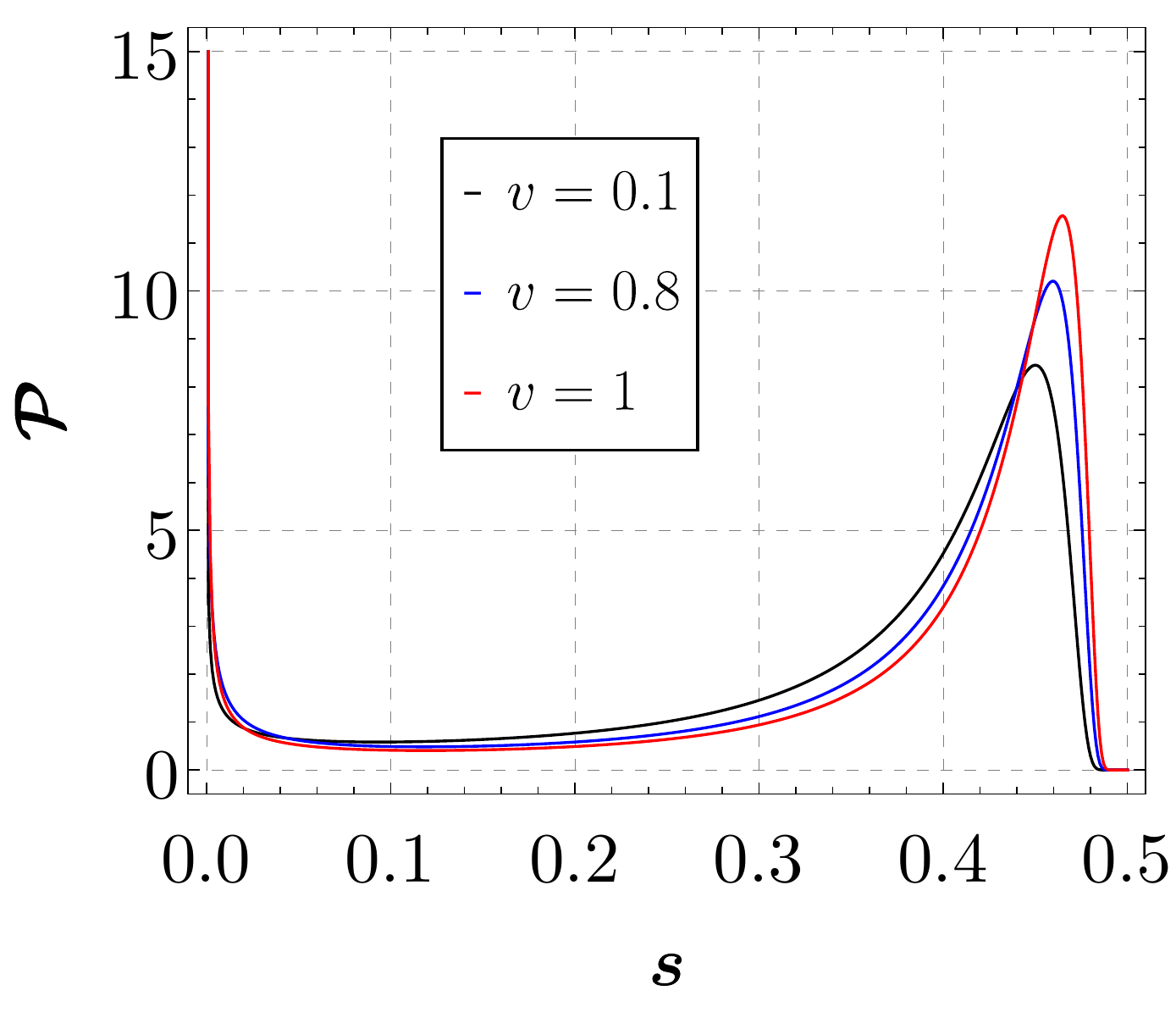} 

    \caption{ Plot of $\calP(s,t)$ for fixed $\lambda=100$, $\Delta=1/4$, and three choices of $v$, to highlight the role of the phase factor in \eqref{PfiniteT}. 
    }
    \label{figFiniteT}
\end{figure}

The computation for the distribution $\calP(s,t)$ is similar to the Brownian SYK. The main difference is that there are more independent terms in $\xi^\dagger_j \xi_j$ than its infinite temperature counterpart, see Appendix~\ref{app: finite T} for details. 
The upshot is that the different terms come with different phases comparing to \eqref{calPint}, and resulting in the following expression for finite temperature size distribution 
\begin{equation}\label{PfiniteT}
    \calP(s,t) = \int_0^{+\infty} h^\R(y) \delta\left( s- \frac{1-\frac{f^\A(\lambda e^{iv\pi/2} y)+f^\A(\lambda e^{-iv\pi/2} y)}{2}}{2}  \right) dy
\end{equation}
which reduces to the infinite temperature result \eqref{calPint} when the Lyapunov exponent is negligible comparing to the temperature, i.e. $v=\varkappa/2\pi T=0$. 
Inserting the explicit form of the functions 
$ h^\R(y)=  y^{2\Delta-1} e^{-y}/\Gamma(2\Delta)$ and $f^\A(x)= (1+x)^{-2\Delta}$ obtained in Ref.~\cite{Gu:2021xaj}, we plot the size distribution with fixed $\lambda= (4N \Delta^2 \cos \frac{\pi v}{2})^{-1}e^{\frac{2\pi v}{\beta}t}$ and different $v$ in Fig.~\ref{figFiniteT}, to emphasize the role of the non-trivial phases $e^{\pm iv \frac{\pi}{2} }$.

\section{Summary and discussions} 
In this paper, we provide a method to determine the size distribution function $\calP(s,t)$ for large $N$ Majorana systems with many-body chaos, extending the existing literature about initial growth to the full time range. 

Remarkably, the notion of operator size has been related to the quantum teleportation via traversable wormholes
\cite{Brown:2019hmk,Nezami:2021yaq,Gao:2019nyj}. Indeed, some of our discussions have counterparts in holography. For example, the integral form of the generating function \eqref{CalSint} is similar to a two point function on the background of a traversable wormhole 
\cite{Maldacena:2017axo}, where variable $y$ corresponds to the momentum and $f^\A(\lambda y)$ represents a two point function on a shifted background parameterized by $y$. 

Furthermore, the size distribution may be generally regarded as a new type of ``order parameter'' that refines the OTOC in diagnosing information scrambling. 
Recently, Refs.~\cite{Schuster:2022bot,Zhang:2022knu} considered the evolution of operator size in open quantum systems.
In particular, \cite{Zhang:2022knu} showed that the size distribution can be used to exhibit dynamical transitions between the scrambling phase and the dissipative phase for systems coupled to a heat bath. 


\section*{Acknowledgment}

We thank Cheng Peng, Xiao-Liang Qi, Zhenbin Yang, and Shunyu Yao for discussions. 
We thank Alexei Kitaev for discussion and collaboration on related projects. 
PZ is partly supported by the Walter Burke Institute for Theoretical Physics at Caltech.
YG is partly supported by the Simons Foundation through the “It from Qubit” program.

\appendix

\section{Numerics for the Brownian SYK model}
\label{app: brownian}

We apply the method developed in Ref.~\cite{Sunderhauf:2019djv} for our numerics presented in Fig.~\ref{figBrownian}~(b). The idea is that the averaged dynamics of the Brownian SYK model can be simulated using bosonic collective modes with a polynomial computational complexity. We refer readers to  Ref.~\cite{Sunderhauf:2019djv} for the conventions and the details of the formalism, and only present the augmentation that is used in our case.

The quantities relevant to the operator size distribution are the OTOCs 
\begin{equation}
\label{boson}
    \mathcal{F}_{(p,n,m)}(t):=\underbrace{\left<\Phi^{(p,n)}(0)^\dagger\Phi^{(p,m)}(t)^\dagger\Phi^{(p,n)}(0)\Phi^{(p,m)}(t)\right>}_{\text{fermion picture}}=\underbrace{\langle\langle LR_{(p,n,m)}|~e^{Ht}~|U_0\rangle\rangle}_{\text{boson picture}}, 
\end{equation}
where the operators $ \Phi^{(p,n)}$, $H$ and the state $|U_0\rangle\rangle$ are defined in Ref.~\cite{Sunderhauf:2019djv}. The only difference is that we have included the additional hermitian conjugation for the first two operators in our formula (cf. Eq.~(3.34) of \cite{Sunderhauf:2019djv}), which leads to the following phase factor in $|LR_{(p,n,m)}\rangle\rangle$:
\begin{equation}\label{eq_LRpnm}
    |LR_{(p,n,m)}\rangle\rangle=\frac{\sqrt{8}^N}{\sqrt{N!}}(-1)^{\frac{(m+n)(m+n+1)}{2}+\frac{(m+p)(m+p-1)}{2}+\frac{(n+p)(n+p-1)}{2}}(b_3^\dagger)^p(b_2^\dagger)^n(b_4^\dagger)^m(b_1^\dagger)^{N-p-n-m}|\Omega\rangle.
\end{equation}


In the main text, we have plotted the moments $\bar{(2s)^m}$ with $m=1,2,3$ for the size distribution. They are related to the function $\mathcal{F}_{(p,n,m)}$ as follows,
\begin{equation}
\begin{aligned}
    \bar{(2s)}&=1+\frac{1}{N} \overline{\text{OTOC}}_1,\\
    \overline{(2s)^2}&=1+\frac{2}{N}\overline{\text{OTOC}}_1+\frac{1}{N^2} \overline{\text{OTOC}}_2,\\
    \overline{(2s)^3}&=1+\frac{3}{N} \overline{\text{OTOC}}_1+\frac{3}{N^2} \overline{\text{OTOC}}_2+\frac{1}{N^3} \overline{\text{OTOC}}_3.
\end{aligned}
\end{equation}
where 
\begin{equation}
\begin{aligned}
    \overline{\text{OTOC}}_1=&(N-1)\mathcal{F}_{(0,1,1)}+\mathcal{F}_{(1,0,1)},\\
    \overline{\text{OTOC}}_2=&N+2(N-1)\mathcal{F}_{(1,1,0)}+(N-1)(N-2)\mathcal{F}_{(0,2,1)},\\
    \overline{\text{OTOC}}_3=&(N-1)(N-2)(N-3)\mathcal{F}_{(0,3,1)}+3(N-1)(N-2)(\mathcal{F}_{(1,2,0)}+\mathcal{F}_{(0,1,1)})\\&
    +3(N-1)(\mathcal{F}_{(0,1,1)}+\mathcal{F}_{(1,0,0)})+(N-1)\mathcal{F}_{(0,1,1)}+\mathcal{F}_{(1,0,0)}.
\end{aligned}
\end{equation}
Numerically simulate \eqref{boson} for the choices of $(p,n,m)$ appear in the above equation at 
$q=4$ and $N=500$, we obtain the data in Fig.~2 (b).

\section{Details of the finite temperature operator size and comparison with Qi-Streicher's definition}
\label{app: finite T}

In this section, we present details of the finite temperature calculation with our definition of size \eqref{eq: defSize}. 
Again, we start with the generating function
\begin{equation}
\label{finiteS}
    \calS(\nu,t)=(c_\beta)^{-1}\langle \rho_\beta^{1/4}\chi_k(t) \rho_\beta^{1/4}|e^{-\frac\nu N (c_\beta)^{-1} \sum_j \xi^\dagger_j  \xi_j}|\rho_\beta^{1/4} \chi_k(t) \rho_\beta^{1/4}\rangle.
\end{equation}
The prefactor $c_\beta=\langle \rho_{\beta}^{1/4} \chi_k(t)  \rho_{\beta}^{1/4} |  \rho_{\beta}^{1/4} \chi_k(t)  \rho_{\beta}^{1/4} \rangle$ in front of the expectation value and also in the exponent 
arises from the normalization in the definition of size \eqref{eq: defSize}.

\begin{figure}[t]
\centering
\begin{tabular}{c@{\hspace{3cm}}c}
\begin{tikzpicture}[scale=0.5,baseline={([yshift=-7pt]current bounding box.center)}]
\draw [->,>=stealth] (-50pt,-57pt) -- (180pt,-57pt) node[right]{\scriptsize  $t$};
\draw [->,>=stealth] (0pt, -135pt) -- (0pt,40pt);
\draw[thick,gray] (0pt,20pt)--(0pt,-20pt);
\draw[thick,far arrow,gray] (0pt,-20pt)--(140pt,-20pt);
\draw[thick,far arrow,gray] (140pt,-24pt)--(0pt,-24pt);
\draw[thick,gray] (0pt,-24pt)--(0pt,-90pt);
\filldraw (140pt,-22pt) circle (2pt) node[right]{\scriptsize$(t+i\beta/4)$};

\draw[thick,far arrow,gray] (0pt,-90pt)--(140pt,-90pt);
\draw[thick,far arrow,gray] (140pt,-94pt)--(0pt,-94pt);
\draw[->,>=stealth, thick,gray] (0pt,-94pt)--(0pt,-130pt); 
\node at (30pt,-130pt) {\scriptsize $-i\beta/2$};
\node at (30pt,20pt) {\scriptsize $i\beta/2$};
\filldraw (140pt,-92pt) circle (2pt) node[right]{\scriptsize$(t-i\beta/4)$};
\filldraw (0pt,-60pt) circle (0pt) node[below right]{\scriptsize $0$};
\filldraw[blue] (0pt,-87pt) circle (2pt) node[left]{\scriptsize$+$};
\filldraw[red] (0pt,-17pt) circle (2pt) node[left]{\scriptsize$+$};
\filldraw[blue] (0pt,-97pt) circle (2pt) node[left]{\scriptsize$-$};
\filldraw[red] (0pt,-27pt) circle (2pt) node[left]{\scriptsize$-$};
\end{tikzpicture}
 &
\begin{tikzpicture}[scale=0.5,baseline={([yshift=-7pt]current bounding box.center)}]
\draw [->,>=stealth] (-50pt,-57pt) -- (180pt,-57pt) node[right]{\scriptsize $t$};
\draw [->,>=stealth] (0pt, -135pt) -- (0pt,40pt);
\draw[thick,gray] (0pt,20pt)--(0pt,-50pt);

\draw[thick,gray,far arrow] (0pt,-50pt)--(140pt,-50pt);
\draw[thick,gray,far arrow] (140pt,-54pt)--(0pt,-54pt);
\draw[thick,gray] (0pt,-54pt)--(0pt,-60pt);
\filldraw (140pt,-52pt) circle (2pt) node[above right]{\scriptsize$(t+i\varepsilon)$};
\draw[thick,gray,far arrow] (0pt,-60pt)--(140pt,-60pt);
\draw[thick,gray,far arrow] (140pt,-64pt)--(0pt,-64pt);
\draw[thick,gray,->,>=stealth] (0pt,-64pt)--(0pt,-120pt);
\filldraw (140pt,-62pt) circle (2pt) node[below right]{\scriptsize$(t-i\varepsilon)$};

\filldraw (0pt,-60pt) circle (0pt) node[below right]{\scriptsize $0$};
\filldraw (0pt,-125pt) circle (0pt) node[above right]{\scriptsize $-i\beta/2$};

\filldraw (0pt,0pt) circle (0pt) node[above right]{\scriptsize $i\beta/2$};

\filldraw[blue] (0pt,-57pt) circle (2pt);
\filldraw[red] (0pt,20pt) circle (2pt);

\end{tikzpicture}
\vspace{3pt}\\
(a) symmetric configuration & (b) Qi-Streicher's definition
\end{tabular}
\caption{The path-integral contour for the generating function of the operator size distribution: (a) the definition in the main text \cite{Gu:2021xaj} and (b) Qi-Streicher's definition.
}
\label{figcontourfinite}
\end{figure}
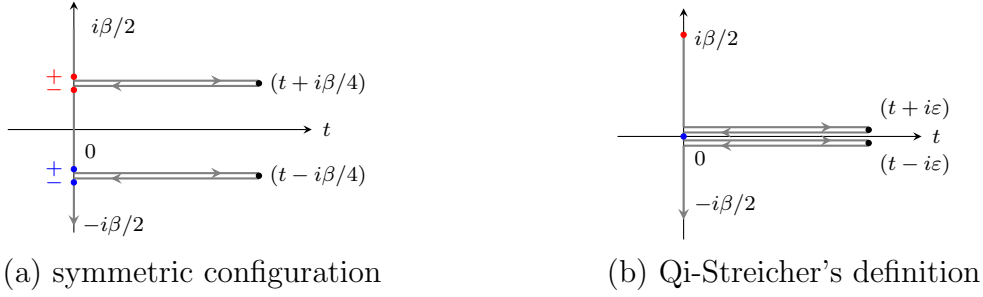

To proceed, we represent \eqref{finiteS} on the double Keldysh contour as 
\begin{equation}\label{calSPIfinite}
    \mathcal{S}(\nu,t)= (c_\beta)^{-1} \langle {\bf\rm T}_c~ \chi_k(t-i\beta/4) e^{\frac{\nu}{N} \sum_j \sum_{\eta_3,\eta_4=\pm}\eta_3\eta_4\frac{\chi_j(-i\beta/4 -i0^{\eta_3} )\chi_j(i\beta/4-i0^{\eta_4})}{4c_\beta} }  \chi_k(t+i\beta/4) \rangle.
\end{equation}
where $i0^\pm= \pm i\varepsilon $ is an infinitesimal shift on the imaginary time, to indicate the positions of the operators as shown Fig.~\ref{figcontourfinite}~(a). 

The four terms in the exponent can be grouped into two classes: $(\eta_3,\eta_4)=(++),(--)$, corresponding to the OTOCs, and $(\eta_3,\eta_4)=(+-),(-+)$, corresponding to the normal-order correlators that can be factor out in the large $N$ limit. Therefore, we have 
\begin{equation}
        \mathcal{S}(\nu,t)
    =(c_\beta)^{-1} e^{-\frac{\nu}{2}}\langle {\bf\rm T}_c~ \chi_k(t-i\beta/4) e^{\frac{\nu}{N} \sum_j \frac{\chi_j(-i(\beta/4+\varepsilon ))\chi_j(i(\beta/4-\varepsilon))
    + \chi_j(-i(\beta/4- \varepsilon ))\chi_j(i(\beta/4+ \varepsilon))
    }{4c_\beta} }  \chi_k(t+i\beta/4)  \rangle.
\end{equation}
Following the same diagrammatic analysis as in the main text, we obtain the following formula 
\begin{equation}
    \mathcal{S}(\nu,t)=\int_0^\infty dy~ h^\text{R}(y) e^{-\nu\left(\frac{1}{2}-\frac{f^\text{A}(\lambda e^{ i\varkappa \beta/4}y)+f^\text{A}(\lambda e^{- i\varkappa \beta/4}y)}{4}\right)},
\end{equation}
where 
\begin{equation}
    h^\R(y) = \frac{y^{2\Delta-1}  e^{-y}}{\Gamma (2\Delta)}, \qquad f^\A(x) = \frac{1}{(1+x)^{2\Delta}} 
\end{equation}
are obtained by restricting the general functions $ h^{\R/\A}(y,\theta)$, $f^{\R/\A}(x,\theta)$ (derived in Ref.~\cite{Gu:2021xaj}, cf. section 6.3) to special case when $\theta=\beta/2$ (we have further simplified the expressions by eliminating common factor $c_\beta$). Inverse Laplace transforming $\mathcal{S}(\nu,t)$ to $\calP(s,t)$, we obtain \eqref{PfiniteT}.

For the rest of this section, we would like to compare our definition of the finite temperature operator size with Qi-Streicher's definition proposed in Ref.~\cite{Qi:2018bje}. More specifically, Qi and Streicher considered the ``size'' of the combination ${\chi}_k(t)\sqrt{\rho_\beta}$ in the sense of 
 the ``length'' of the operator basis, i.e. measured by $\sum_j c^\dagger_j  c_j$ as in the infinite temperature case. 
 In this definition, the generating function is given as 
\begin{equation}
    \calS(\nu,t)=\langle {\chi}_k(t)\sqrt{\rho_\beta}|e^{-\frac\nu N \sum_j c^\dagger_j  c_j}|{\chi}_k(t)\sqrt{\rho_\beta}\rangle.
\end{equation}

Now, we apply our method to compute the generating function under Qi-Streicher's definition. 
There are two key differences: (1) our definition involves two OTOC configurations while Qi-Streicher's definition only involves one; 
(2) as shown in Fig.~\ref{figcontourfinite}~(b), the two Keldysh contours are placed in different locations, which invokes more general functions $ h^{\R/\A}(y,\theta)$, $f^{\R/\A}(x,\theta)$. 
Nevertheless, it still yields an explicit formula in the large-$q$ SYK model 
\begin{equation}
    \mathcal{S}(\nu,t)=\int_0^\infty dy~ \tilde{h}^\text{R}(y) e^{-\frac{\nu}{2}\left(1-c_\beta f^\text{A}(\lambda y)\right)},
\end{equation}
where a new function 
\begin{equation}
    \tilde{h}^\text{R}(y)=c_\beta\frac{ y^{2\Delta-1}}{\Gamma(2\Delta)}e^{-c_\beta^{1/(2\Delta)} y }
\end{equation}
is needed. Inserting the functions $\tilde{h}^\R$ and $f^\A$ (defined before), we obtain the following explicit form for the size distribution via inverse Laplace transform 
\begin{equation}\label{eq_Qisize}
    \mathcal{P}(s,t)=\int_0^\infty dy~ \tilde{h}^\text{R}(y) \delta\left(s-\frac{1-c_\beta f^\text{A}(\lambda y)}{2}\right) =\frac{2c_\beta^q  \sigma_S ^{2 \Delta -1} (1-2 s)^{-\frac{2 \Delta +1}{2 \Delta
   }}}{\lambda  \Gamma (2 \Delta +1)e^{\sigma_S }},
\end{equation}
with $\lambda\sigma_S={c_\beta^{1/\Delta}(1-2 s)^{-1/2\Delta }-c_\beta^{1/2\Delta}}$. From the first step, 
where $f_A \in (0,1) $, we can see that 
the size in Qi-Streicher definition ranges from ${1-c_\beta}/{2}$ to $1/2$, consist with the result in \cite{Qi:2018bje}.

Eq. \eqref{eq_Qisize} also has implications on the relation between Qi-Streicher size and complexity. At short time, we can expand $f^\text{A}$ in terms of $\lambda \ll1$. This implies the time dependent part of size $s$ is proportional to the null momentum of the particle. In the long-time limit, the result shows the operator size is determined by the correlation between two sides on the perturbed geometry. Assuming the complexity-volume conjecture in \cite{Stanford:2014jda}, this implies in AdS$_2$ we have $-\log ({1/2-s})\propto \mathcal{C}$, where $\mathcal{C}$ is the computational complexity of the quantum state.

\bibliography{Draft_single.bbl}

\end{document}